\documentstyle[11pt]{article}
\let\LARGE=\Large
\let\Large=\large
\let\large=\normalsize
\newcommand{\be}{\begin{equation}} 
\newcommand{\ee}{ \end{equation}}
\newcommand{\ba}{\begin{array}}
\newcommand{\ea}{\end{array}}
\newcommand{\bea}{\begin{eqnarray}}
\newcommand{\eea}{\end{eqnarray}}
\newcommand{\NP}[3]{{\em Nucl. Phys.}{ \bf B#1#2#3}}

\newcommand{\ft}[2]{{\textstyle\frac{#1}{#2}}}

\newcommand{\ov}{\overline}
\def\beq{\begin{equation}}
\def\eeq{\end{equation}}
\def\beqa{\begin{eqnarray}}
\def\eeqa{\end{eqnarray}}

\newcommand{\eqn}[1]{(\ref{#1})}

\renewcommand{\d}{\delta}
\newcommand{\pa}{\partial}

\newcommand{\m}{\mu}

\newcommand{\n}{\nu}

\newcommand{\s}{\sigma}

\newcommand{\la}{\langle}
\newcommand{\ra}{\rangle}

\newcommand{\mscr}[1]{\mbox{\scriptsize #1}}
\newcommand{\fscr}[1]{\mbox{\scriptsize \bf #1}}

\begin{document}
\begin{titlepage}
\begin{center}
\hfill AEI-111\\[1mm]
\hfill THU-99/11\\
\hfill {\tt hep-th/9906094}\\

\vskip 3cm

{ \LARGE \bf Macroscopic entropy formulae and non-holomorphic corrections 
for supersymmetric black holes}

\vskip .3in

{\bf Gabriel Lopes Cardoso$^{1}$\footnote{\mbox{
\tt 
cardoso@phys.uu.nl}}, 
Bernard de Wit$^{2}$\footnote{\mbox{
\tt 
bdewit@phys.uu.nl}} 
and Thomas Mohaupt$^3$\footnote{\mbox{
\tt 
mohaupt@hera1.physik.uni-halle.de}}}
\\

\vskip 1cm

{\em 


\centerline{$^{1,2}$Institute for Theoretical Physics, Utrecht University,
3508 TA Utrecht, The Netherlands}
\vspace{1.5mm}
\centerline{$^2$Max-Planck-Institut f\"ur Gravitationsphysik,
Albert-Einstein-Institut,} 
\centerline{Haus 5, Am M\"uhlenberg, D-14476 Golm, Germany}
\vspace{1.5mm}
\centerline{$^3$Martin-Luther-Universit\"at Halle-Wittenberg, 
Fachbereich Physik,
D-06099 Halle, Germany}    }

\vskip .1in
\end{center}

\vskip .2in

\begin{center} {\bf ABSTRACT } \end{center}
In four-dimensional $N=2$ compactifications of string theory or
M-theory, modifications of the Bekenstein-Hawking area law for
black hole entropy in the presence of higher-derivative interactions
are crucial for finding 
agreement between the macroscopic entropy obtained from
supergravity and subleading corrections to
the microscopic entropy obtained via state counting.
Here we compute the modifications to the area law for various classes
of black holes, such as heterotic black holes, stemming from certain
higher-derivative gravitational Wilsonian coupling functions. 
We consider the
extension to heterotic $N=4$ supersymmetric black holes 
and their type-II duals 
and we discuss its implications for the corresponding micro-state counting.
In the effective
field theory approach the Wilsonian coupling functions are known to
receive non-holomorphic 
corrections.  We discuss how to incorporate such corrections
into macroscopic entropy formulae so as to render them invariant under
duality transformations, and we give a concrete example thereof.


\vfill

June 1999\\
\end{titlepage}

\section{Introduction}
In the presence of higher-derivative interactions, 
modifications of the Bekenstein-Hawking area law for
the macroscopic entropy of a 
black hole are crucial for finding 
agreement with microscopic entropy calculations based on 
state counting.  The latter involve corrections that are subleading
in the limit of large charges, and these corrections are 
captured on the macroscopic side by the modifications of 
the area law that ensure the validity of the first law of
black hole mechanics in the presence of higher-derivative
interactions.
For supersymmetric black hole solutions arising in
four-dimensional $N=2$ supersymmetric effective field theories
with terms quadratic in the Riemann tensor, they were determined
in \cite{CDWM}, where it was shown that, for a particular class of
black holes arising in compactifications of M-theory and 
type-IIA string theory, the modified macroscopic entropy exactly matches
the microscopic entropy computed in \cite{MSW,Vafa}. 

The four-dimensional supersymmetric black hole solutions considered in
\cite{CDWM} are static, 
rotationally symmetric solitonic interpolations between 
two $N=2$ supersymmetric groundstates: flat Minkowski spacetime at spatial 
infinity and Bertotti-Robinson spacetime at the horizon 
\cite{Gibbons}.  The crucial ingredients of the work reported in
\cite{CDWM} are as follows.  One departs from the
Bekenstein-Hawking area formula and adopts 
Wald's proposal for the entropy based on a Noether charge 
associated with an isometry evaluated at the corresponding  
Killing horizon \cite{Wald,cdwm}.  The computation of the macroscopic
entropy is based on the effective $N=2$ 
Wilsonian action describing the couplings of abelian vector
multiplets to $N=2$ supergravity in the presence of a certain class of
terms quadratic in the Riemann tensor.  This Wilsonian action is
determined from a holomorphic, homogeneous  
function $F(Y, \Upsilon)$, where the $Y^I$ denote the (rescaled) 
complex scalar fields which reside in the abelian vector multiplets,
and where $\Upsilon$ denotes the (rescaled) square of the (auxiliary)
anti-selfdual Lorentz tensor field $T^{ab \, ij}$ which resides in the
Weyl multiplet. 
The macroscopic entropy of a  static, supersymmetric black hole 
computed from the effective Wilsonian Lagrangian 
is then given by \cite{CDWM}
\beqa
{\cal  S}_{\rm macro} = \pi \Big[ |Z|^2 +4 \, {\rm Im} \Big( 
\Upsilon \,F_{\Upsilon}
(Y,\Upsilon) \Big)
\Big] \;\;\; \mbox{with} \;\;\; \Upsilon = -64 \;\;\;.
\label{entropiay}
\eeqa
The first term in the entropy formula, 
$|Z|^2= p^I F_I( Y , \Upsilon) - q_I Y^I$,
coincides with the
Bekenstein-Hawking area contribution.  The values of the
scalar fields $Y^I$ on the horizon are, when assuming fixed-point behaviour
\cite{FKS,Moore}, determined in terms of the magnetic and electric
charges $(p^I, q_I)$ carried by the black hole.  Their precise value
is determined from a set of equations, called the stabilization
equations, which take the following form \cite{BCDWLMS}:
\beqa
Y^I-\bar Y{}^I= i\, p^I \;\;,\qquad 
F_I(Y,\Upsilon) -\bar F_I(\bar Y, \bar \Upsilon) = i\, q_I 
\;\;\;.
\label{staby}
\eeqa
The entropy is thus entirely determined in terms of the charges carried by
the extremal black hole. 

The above  equations can usually not be solved in explicit
form. Often they can be solved by iteration,
order-by-order in $\Upsilon$. Let $Q$ denote a generic electric or
magnetic charge. Then the homogeneity of the function $F$ implies that
the entropy can be expanded in powers of $Q$, according to
\be
{\cal S}_{\rm macro} 
= \pi \sum_{g=0}^\infty  a_g\, Q^{2-2g}\,, \label{expansion}
\ee
with constant coefficients $a_g$. These coefficients follow from the
coefficient functions in the power expansion for the holomorphic
function $F$, which takes the form  $F(Y, \Upsilon)=
\sum_{g\geq 0} F^{(g)}(Y) \, \Upsilon^g$. 

In this paper we present a  variety of applications of the 
entropy formula \eqn{entropiay}. 
In particular we consider $N=2$ heterotic black holes,
whose macroscopic entropy formula can be generalized to the case of $N=4$
supersymmetric black holes. In order to appreciate the consequences
of our results we will return to the counting of micro-states in the
dual type-II  or M-theory version of these black holes, following the
same approach as in \cite{MSW,Vafa}. In this analysis certain  special
features of $K3\times T^2$ emerge which are not present for generic Calabi-Yau
threefolds, implying that the micro-state counting in the former case 
is more subtle than in the latter case.  Based on this analysis we propose
a modification of the state counting for $K3\times T^2$ which is consistent
with the macroscopic result. 

Several comments are in order here.  First, as mentioned above, the entropy
formula (\ref{entropiay}) is based on the Wilsonian Lagrangian
approach.  In this approach, 
the coupling function multiplying the square of the Weyl tensor  
$C_{\mu \nu \rho \sigma}$ is denoted by 
${\rm Im} \,F_{\Upsilon}$.  It is customary to assume that $F_{\Upsilon}$
possesses a power expansion in $\Upsilon$, i.e.,  $F_{\Upsilon} = 
\sum_{g \geq 1} F^{(g)} (Y) \Upsilon^{g-1}$.  As pointed out recently
in \cite{GV}, this is not necessarily the case due to the 
fact that the $F_{\Upsilon}$ is non-analytic at $\Upsilon=0$.
Second, the stabilization equations
(\ref{staby}) are constructed out of the two symplectic
vectors $(p^I, q_I)$ and $(Y^I, F_I (Y, \Upsilon))$.  It is important
to note that under symplectic transformations the $Y^I$ will in general
transform into $\Upsilon$-dependent terms.  Thus, under the subset of
duality transformations the moduli fields $z^A = Y^A/Y^0$ will, in
the presence of $C^2_{\m\n\rho\s}$-terms (that is
$F_\Upsilon\not=0$), transform
into $\Upsilon$-dependent quantities! 
Finally, we recall that in theories with massless fields the physical
effective couplings are different from the Wilsonian couplings, which
presuppose the presence of an infrared cut-off. While the physical
couplings have different analyticity properties, 
they fully capture the
physics and its underlying symmetries. Thus, in order to arrive
at entropy formulae that are manifestly duality invariant, it is necessary
to modify both (\ref{entropiay}) and (\ref{staby}), which were derived in 
a Wilsonian approach, by taking into account non-holomorphic corrections
to the physical effective couplings. In the context of string theory
one can actually determine these non-holomorphic terms, because the
massless states are included in the string loops. Here we choose to
incorporate the non-holomorphic 
corrections by imposing the correct symmetry behaviour. Therefore 
we include certain non-holomorphic corrections to the
$\Upsilon$-dependent part of the function $F$,
or equivalently, to $F_{\Upsilon} (Y,{\bar Y}, \Upsilon)$. Here the
corresponding  
$F(Y, {\bar Y}, \Upsilon)$ is homogenous of degree two
in $Y$ and of degree zero in ${\bar Y}$.  The corrections to 
$F_{\Upsilon}$ should be such that the associated symplectic vector
$(Y^I, F_I (Y, \bar Y, \Upsilon))$ transforms in the same way as $(p^I,q_I)$
under symplectic transformations associated with electric/magnetic duality.  
The associated stabilization equations now read
\beqa
Y^I-\bar Y{}^I= i\, p^I \;\;,\qquad 
F_I(Y, {\bar Y},\Upsilon) -\bar F_I( Y, \bar Y, \bar \Upsilon) =
i\,q_I   \;\;. \label{nhstaby}
\eeqa
Provided that the duality transformations on the $Y^I$ induce the
corresponding symplectic transformations on the $F_I$, the quantity
$|Z|^2 = p^I 
F_I( Y , {\bar Y}, \Upsilon)  - q_I Y^I$ is then invariant under
duality transformations.   
However, in order to arrive at a  duality invariant expression
for the entropy of an $N=2$ black hole,  it will not suffice to 
only take into account the above modification of $F_{\Upsilon}$ and, hence,
of $|Z|^2$.  As it turns out, also the second term associated with the
deviation from the area law requires a non-holomorphic correction. Hence
we are forced to use the following expression for the entropy:
\beqa
{\cal  S}_{\rm macro} 
= \pi \Big[ p^I F_I(Y, {\bar Y}, \Upsilon) - q_I Y^I 
 +4 \, {\rm Im} \Big(\Upsilon\, F_{\Upsilon} (Y, {\bar Y}, \Upsilon) + 
\Delta (Y, {\bar Y}, \Upsilon) 
\Big) \Big]
\; , \;
{\rm with} \; \; \Upsilon = -64 \;,
\label{nhentropiay}
\eeqa
where $\Delta (Y, \bar Y, \Upsilon)$ denotes an appropriate quantity
whose role is to render ${\rm Im} \, (\Upsilon\, F_{\Upsilon} + \Delta
)$ duality invariant.  In general it should be complicated
to determine the explicit form of the 
non-holomorphic corrections entering both in (\ref{nhstaby})
and (\ref{nhentropiay}) due to the fact, mentioned above, that the moduli
fields $z^A$ will transform into $\Upsilon$-dependent terms under
duality transformations.  This is to be contrasted with the usual approach
for determining non-holomorphic corrections to physical couplings
in string theory \cite{Bershadskyetal,antganartay,deWit}, in which
each of the gravitational coupling functions in the expansion of the
holomorphic function $F$ in powers of $\Upsilon$ is required to
transform as a modular function of the appropriate weight.  This is achieved by
taking into account non-holomorphic corrections to these
coupling functions, which are governed by a set of recursive
holomorphic anomaly equations.  In the case at hand, however, 
not only does 
$ F ( Y , {\bar Y}, \Upsilon )$ and therefore also the
symplectic vector $(Y^I, F_I ( Y , {\bar Y}, \Upsilon))$ 
and $|Z|^2$ receive non-holomorphic corrections, but there 
are additional non-holomorphic contributions encoded in $\Delta$. 
While this is an important feature of the above mechanism for
incorporating non-holomorphic corrections into macroscopic entropy
formulae, its real significance is not yet clear to us.

The structure of this paper is as follows.  In section 2
we compute the macroscopic 
entropy of a generic tree-level black hole in a 
heterotic $N=2$ compactification in the presence of a tree-level
$C^2_{\mu \nu \rho \sigma}$-term. Subsequently we consider its
extension to $N=4$ supersymmetric black holes. 
In section 3 we then reconsider the $N=2$ and $N=4$ counting of
micro-states in the dual type-II and M-theory formulations of the
black holes. We show that the expression
for the macroscopic $N=4$ entropy agrees with a microscopic
entropy computation along the lines of \cite{MSW,Vafa} 
for a class of black holes occuring
in dual type-II compactifications on $K3 \times T^2$, provided
that the micro-state counting for $K3 \times T^2$ is suitably modified,
as we discuss.
In section 4 we turn to the issue of non-holomorphic corrections
to the macroscopic entropy.  We discuss a heterotic-like $N=2$
model for which we are able to determine the non-holomorphic
corrections to the entropy needed for obtaining a 
strong-weak coupling
duality invariant expression for the entropy.  
And finally, in section 5, we use the recent results of \cite{GV} to
compute corrections to the macroscopic
entropy of certain type-IIA black holes stemming from $T^{ab ij}$-dependent
higher-derivative interaction terms in the effective action.

\section{Heterotic tree-level black holes \label{treehetb}}
\setcounter{equation}{0}
In the presence of an $(S+\bar S) \, C^2_{\mu \nu \rho\sigma}$-term in the
effective Wilsonian action, where $S$ is the dilaton field, the
tree-level holomorphic function $F(Y,\Upsilon)$ associated with a 
heterotic $N=2$ compactification on $K3 \times T^2$ is  given by
\bea
F(Y,\Upsilon) = - \frac{Y^1 Y^a \eta_{ab} 
Y^b}{Y^0} + c_1 \; \frac{Y^1}{Y^0} \; \Upsilon \;\;.
\label{hetprep}
\eea
Here 
\bea
 Y^a \eta_{ab} Y^b = Y^2 Y^3 - \sum_{a=4}^{n} (Y^a)^2 \;,\qquad 
a=2, \ldots, n \;,
\eea
with real constants $\eta_{ab}$ and 
$c_1$. In a type-II dual picture, these constants can be related to
geometrical properties of the $K3$ fibration of the corresponding
Calabi-Yau space. 
The dilaton field $S$ is defined by $S =  -i Y^1/Y^0$.

Using \eqn{staby} with $I=a$ one readily proves that the $Y^a$ are
given by 
\be
Y^a = {1\over S+\bar S} \Big[ -\ft12 \eta^{ab} q_b +i\bar S \,p^a
\Big]\, , 
\ee
where $\eta^{ab}\,\eta_{bc} = \delta^a_{\,c}$. 
Using the same equations once more one establishes
\bea
\vert Z\vert^2 &=&  p^I F_I  - q_I Y^I\nonumber\\[2mm]
&=& i (\bar Y^I F_I - Y^I \bar F_I)  \nonumber \\
&=& (S+\bar S) \Big(\,\bar Y^a \eta_{ab} Y^b + {\bar Y^0\over Y^0}
\Big[- Y^a \eta_{ab} Y^b   + c_1\, \Upsilon \Big]
+ \mbox{ h.c.}\Big) \,,
\eea
as well as
\bea
q_1\,p^0 &=& - (Y^0- \bar Y^0 ) (F_1 - \bar F_1) \nonumber\\
&=& \Big({\bar Y^0 \over Y^0}-1\Big)   \Big[- Y^a
\eta_{ab} Y^b   + c_1\, \Upsilon  \Big]
+ \mbox{ h.c.}\,.
\eea
Combining these two equations and substituting the values for the
$Y^a$ yields 
\bea
|Z|^2  = (S + {\bar S}) \Big( \ft{1}{2} \, \langle N,N \rangle
+ (c_1 \,\Upsilon + \mbox{ h.c.}) \, \Big)
\;,
\label{hzz}
\eea
where $\langle N,N \rangle 
= 2( N^0 N^1 + N^a \eta_{ab} N^b ) = 
2 (  p^0 q_1 + p^a\eta_{ab} p^b)$. The vector $N^I$ of magnetic
charges (in a certain duality basis) is defined
by $N^I = (p^0, q_1, p^2, p^3, \dots, p^n)$, and  transforms linearly
under the target-space duality 
group SO$(2,n-1)$. The second independent vector of electric charges,
which also 
transforms linearly under the target-space duality group, will be
needed shortly and
equals $M_I = ( q_0, - p^1, q_2, q_3, \dots, q_n)$. 

{From} \eqn{hzz} we can now determine the entropy of a generic heterotic
black hole, by using \eqn{entropiay}. The second term in the formula
doubles the coefficient of the second term in \eqn{hzz} and we find
\bea
{\cal S}_{\rm macro} = \ft{1}{2}\, \pi \, (S + {\bar S})
\, \Big(\langle N,N \rangle - 512 \, c_1 \,   \Big)
\;. 
\label{entros}
\eea
For $c_1=0$ this expression coincides with the one given in 
\cite{BCDKLM}, where it was also shown that 
in the presence of
string-loop corrections the entropy takes the same form as
the tree-level entropy, but with
the heterotic tree-level coupling constant $S + \bar S$ replaced
by the perturbative coupling constant.

Now we continue and find the equations that govern the value of the
dilaton field in terms of the charges. Just as above we use the
equations \eqn{staby} to write down expressions for the combinations 
$S\bar S \, q_1\,p^0 + q_0\,p^1$ and $i(\bar S -S) \, q_1\,p^0+
q_1\,p^1 - q_0 \,p^0$, which do not explicitly depend on $Y^0$.
This leads to two equations for $S$:
\bea
S\bar S \,\langle N,N\rangle &=&  \langle M,M\rangle  -2(S + {\bar
S}) ( c_1\,\Upsilon \,S + \mbox{ h.c.})\,, \nonumber\\
(S- {\bar S}) \,\langle N,N\rangle &=& 2i\, M\cdot N + 2 \,(S + {\bar S})
\,(c_1\,\Upsilon - \mbox{ h.c.} )\,.
\eea
At the horizon this leads to the following results for 
the dilaton and for the entropy formula, 
\bea
S  &=&  
\sqrt{ {\langle M,M \rangle \langle N,N \rangle - ( M \cdot N )^2}\over
{\langle N,N \rangle\,(\langle N,N \rangle  - 512 \,c_1) } } \; + \; i \;
\frac{ M \cdot N }{ \langle N,N \rangle}
\;\;\;, \nonumber\\
{\cal S}_{\rm macro} &=& \pi \; \sqrt{ \langle M,M \rangle 
 \langle N,N \rangle  - ( M \cdot N )^2}
\;  \sqrt{1 - \frac{512 \,c_1 \,}{\langle N,N \rangle}}
\;\;\;,
\label{entrofin}
\eea
where we recall the three target-space duality invariant combinations
of the charges \cite{clm},
\bea
\langle M,M \rangle 
&=& 2 \Big(M_0 M_1 + \ft{1}{4} M_a \eta^{ab} M_b\Big) = 
2\Big( - q_0 p^1 + \ft{1}{4}
 q_a \eta^{ab} q_b \Big) \;\;,
\nonumber\\
\langle N,N \rangle &=& 2\Big( N^0 N^1 + N^a \eta_{ab} N^b \Big) = 
2 \Big(  p^0 q_1 +  p^a \eta_{ab} p^b \Big) \;\;,
\nonumber\\
M \cdot N &=& M_I N^I = q_0 p^0 - q_1 p^1 + q_2 p^2 + \cdots + q_{n}
p^{n} \;\;.
\label{invcomb}
\eea
A significant feature of these results is that they are manifestly
invariant under the continuous SO$(2,n-1)$ target-space duality
transformations, which are valid in the tree approximation. We analyze
these transformations as well as the discrete $S$-duality
transformations in section \ref{secanom}. 

The above entropy formula can be immediately generalized 
to the case of a heterotic 
$N=4$ compactification with a tree-level term proportional to 
$({\rm Re} S) \,C^2_{\mu \nu \rho \sigma}$ turned
on.  This can be done by replacing the SO$(2,n-1)$ bilinear
combinations of the charges by the appropriate SO$(6,22)$ bilinears in
(\ref{entrofin}). Here we remind the reader that the $N=4$
electric/magnetic charges are associated with six 
graviphotons of pure supergravity and 22 vector multiplets, whose
$N=2$ decomposition is as 
follows. One graviphoton belongs to the $N=2$ graviton multiplet, four
graviphotons belong to the two $N=2$ gravitini multiplets (which
cannot be incorporated in the context of the $N=2$ theory discussed
above), and one
graviphoton belongs to an $N=2$ vector multiplet. The additional 22 $N=4$
matter multiplets each 
decompose into an $N=2$ vector and an $N=2$ hypermultiplet, so that
altogether one is dealing with 67 complex moduli. In the $N=2$
truncation one thus has $n=23$ and must suppress 4 of the 28 electric
and 4 of the 28 
magnetic charges. Of course, the $N=2$ truncation does not constitute
by itself a heterotic $N=2$ string compactification, which has only $n=19$ and
which is described by different Wilsonian couplings functions beyond
the tree approximation. 

The $N=4$ supersymmetric  heterotic models have dual
realizations as type-II string theory 
compactified on $K3 \times T^2$ and as eleven-dimensional M-theory
compactified on $K3 \times T^2 \times S^1$. The $N=2$ models have dual
realizations as compactifications of the same theories but now on a
generic Calabi-Yau manifold. 
In the next section we will discuss certain aspects of the
state counting for these dual realizations.

\section{State counting for type-II and M-theory black holes}
\setcounter{equation}{0}
Motivated by the possible extension of the heterotic $N=2$ black hole
entropy formula to the corresponding case of $N=4$ supersymmetric black holes,
we now return to the counting of micro-states 
for the dual type-II and M-theory realizations of the
heterotic models discussed above. If one restricts the
four-dimensional charges in an appropriate way
(we take $p^{0}=q_{A}=0$ for convenience, but the discussion should
also apply to the more general case $p^0=0$ by replacing
$q_0$ by $\hat{q}_0$ as in \cite{MSW}), 
then the black hole
can be given a ten- or eleven-dimensional interpretation, as follows. 
In the type-IIA picture one wraps a D4-brane on a holomorphic 
four-cycle $P$ in either a generic Calabi-Yau threefold or in $K3
\times T^2$ and considers a bound state with  
$|q_0|$ D0-branes. This will then lead to a four-dimensional black
hole with $N=2$ or $N=4$ supersymmetry, respectively. In the M-theory
picture one wraps an M5-brane 
on the same cycle $P$ and obtains a five-dimensional black string which is
then wrapped
around the M-theory circle. Adding $|q_0|$ quanta of lightlike 
momentum along this string yields a four-dimensional black hole
with a finite event horizon.

In both pictures a microscopic entropy can be computed along the
lines of \cite{MSW,Vafa}. 
The approach of reference \cite{MSW} is based on analyzing the 
two-dimensional $\sigma$-model that describes the massless
excitations of an M5-brane wrapped on $P$. 
Then, using Cardy's formula, the asymptotic density
of states of a two-dimensional conformal field theory in
a finite volume is given by 
\be
d(|q_{0}|,c_{L}) \approx \exp({\cal S}_{\mscr{micro}}) \approx 
\exp\Big( 2 \pi \sqrt{ \ft16{|q_{0}| c_{L}} } \,\Big)\;,
\label{cardmicro}
\ee
where $|q_{0}|$ is large and counts quanta of lightlike, left-moving
momentum, and $c_{L}$ is the central charge for the left-moving sector.
On the other hand in reference \cite{Vafa} the asymptotic growth of
the dimension of the moduli space of a bound state
of a wrapped D4-brane with $|q_{0}|$ D0-branes was computed.
The explicit formulae given in \cite{MSW,Vafa} assume certain
properties of the four-cycle $P$, and there it was mentioned that in
the case of 
$K3 \times T^2$ additional subleading terms are to be
expected. 

The holonomy group of $K3 \times T^2$ is ${\rm SU}(2)$, 
unlike the holonomy group of a generic Calabi-Yau
threefold which is equal to SU(3). This difference reflects itself
in the number of residual supersymmetries and in the 
different Hodge diamonds which we will discuss below.  
We will apply the methods of \cite{MSW,Vafa} in parallel to both 
a generic Calabi-Yau threefold and to $K3\times T^2$.
As it will turn out, in the case of $K3 \times T^2$
certain complications arise which indicate that the associated
state counting has to be modified in order to achieve consistency with
the macroscopic analysis based on $N=2$ supergravity.

Let us first recall some geometric properties of Calabi-Yau threefolds 
\cite{GriHar,BarPetVen,Hue,Can}, starting with the homology and
cohomology of a Calabi-Yau threefold $X$, which is either 
$K3 \times T^2$ or a generic one and which corresponds to an $N=4$ or $N=2$
supersymmetric compactification, respectively. In the case of a
generic Calabi-Yau manifold the only undetermined 
Hodge numbers are $h_{2,1}(X)$ and $h_{1,1}(X)$. In addition one
has non-vanishing $h_{0,0}(X)=h_{3,0}(X)=1$, whereas all other independent
Hodge numbers vanish. 

The Hodge numbers of $K3 \times T^2$ are easily obtained from those
of $K3$ and $T^2$. For later convenience let us derive them in terms
of cohomology, i.e. by listing harmonic $(p,q)$-forms. The harmonic
forms on $T^2$ can be built out of the constant function, the
$(1,0)$-form $dz$ and the $(0,1)$-form $d\bar{z}$, where $z$ is a 
holomorphic coordinate. On $K3$ one has the constant function, 
no one-forms and 22 two-forms $\omega_a$, $a=2,\ldots,23$, 
which can be decomposed into one $(2,0)$-form $\omega$, one
$(0,2)$-form $\bar{\omega}$ and twenty $(1,1)$-forms.
The reason for leaving out the index $a=1$ 
will become clear shortly. The
harmonic $(p,q)$-forms on $K3 \times T^2$ are readily found by taking
appropriate products. 
The resulting independent Hodge numbers for $K3\times T^2$ are:
$h_{0,0}= h_{1,0} = h_{2,0}= h_{3,0} = 1$ and 
$h_{1,1} = h_{2,1} = 21$. The two-forms on $K3\times T^2$ will be
denoted by $\omega_A$ with $A= 1,2,\ldots,23$, where
$\omega_1 \propto  dz\wedge d\bar z$ and the $\omega_a$ are the $K3$
forms introduced above. 

The triple intersection
numbers $C_{ABC}$ of a Calabi-Yau threefold are defined by
\be
C_{ABC} = \int_X \omega_A \wedge \omega_B \wedge \omega_C\;.
\ee
As we will be dealing with the integer-valued cohomology group
$H^2(X,{\bf Z})$, the basic two-forms are normalized such that the
intersection numbers are integers. 
For $X=K3 \times T^2$ the appropriate normalization is based on 
\be
\int_{K3} \omega \wedge \ov{\omega} = 1 \,, \qquad
\int_{T^2} \omega_1 = 1\, \quad\mbox{and} \quad
\int_{K3} \omega_{a} \wedge \omega_{b} = C_{ab} \;,
\ee
where $C_{ab}$ is the intersection form of two-cycles in
$K3$. 
The non-vanishing components of the triple intersection form
$C_{ABC}$ of $K3 \times T^2$ are therefore
\be
C_{ab1} = C_{ab}\;.
\ee
In order to discuss the wrapping of D4- or M5-branes on four-cycles, 
we turn to the homology. We recall that
Poincar\'e duality provides a natural relation
between $i$-forms $\omega$ and $2k-i$ cycles $\Omega$ for every real
$2k$-fold $X$ by
\be
\int_{\Omega} \phi = \int_X \omega \wedge \phi\;,
\ee
for all $(2k-i)$-forms $\phi$.
This defines an isomorphism between $H^i(X,{\bf R})$ and
$H_{2k-i}(X,{\bf R})$. In the particular  
case of Calabi-Yau threefolds, four-cycles are dual to two-forms. We can
therefore fix a basis for $H_4(X,{\bf R})$ by taking the four-cycles
dual to a basis of $H^2(X,{\bf R})$. This basis will be denoted by
$\Sigma_A$, where $A=1, \ldots,,b_2(X)$ and where 
$b_2(X) = \dim_{\fscr{R}}H^2(X,{\bf R}) 
=  h_{2,0}(X) + h_{1,1}(X) + h_{0,2}(X)$ 
is the second Betti number. In the case of $X=K3 \times T^2$ 
we have $b_2=23$ and the cycles $\Sigma_A$ with $A\not=1$ are
products of the basic two-cycles of $K3$ with 
$T^2$, whereas $\Sigma_1$ corresponds to $K3$. The four-cycle $P$ on
which the D4- or M5-brane is wrapped can thus be expanded in a
homology basis $\Sigma_A$, 
\be 
P= p^A\,\Sigma_A \,.
\ee 
The integers $p^A$ count how many times the D4- or M5-brane
is wrapped around the corresponding cycle, so that we are actually
dealing with integer-valued 
cohomology and homology groups, $H^2(X,{\bf Z})$ and $H_4(X,{\bf Z})$.
In four-dimensional terms the $p^A$ are magnetic charges, which are
quantized according to Dirac's rule.

By Poincar\'e duality the triple intersection numbers also count
the numbers (weighted with orientation) of intersections of generic
four-cycles in a given homology class.  The triple intersection for a
four-cycle is therefore
\be
P\cdot P \cdot P = C_{ABC}\, p^Ap^Bp^C\,.
\ee
For $K3 \times T^2$ the non-vanishing triple intersections of
four-cycles are
\be
\Sigma_a \cdot \Sigma_b \cdot \Sigma_1 = C_{ab}\;,
\ee
so that the triple intersection of a four-cycle
$P$ in the class $p^a\, \Sigma_a + p^1\, \Sigma_1$ is given by
\be
P \cdot P \cdot P = 3\, C_{ab}\,  p^a p^b p^1\;.
\ee
Another quantity that will appear in the microscopic entropy formula
is the second Chern class $c_2(X)$ of $X$ evaluated on the four-cycle $P$.
Recall that the second Chern class is a four-form which defines 
a cohomology class in $H^4(X,{\bf Z})$. Hence there is a 
dual two-cycle in $H_{2}(X,{\bf Z})$
which we will denote be $C_2$. 
Since $X$ is a Calabi-Yau manifold we can choose a 
basis $\Sigma^{A}$ of $H_{2}(X,{\bf Z})$ which is dual
to the basis $\Sigma_A$  of $H_4(X,{\bf Z})$, 
i.e. $\Sigma^A \cdot \Sigma_B = \delta^A_B$. 
Expanding $C_2 = c_{2A} \Sigma^A$ we find 
\be
C_2 \cdot P = \int_P c_2(X) = c_{2A} \,p^A\;.
\ee
For $K3 \times T^2$ the total Chern class  can be computed 
from the total Chern classes of $K3$ and $T^2$ by the Whitney
product formula:
\be
c(K3 \times T^2) = c(K3) \wedge c(T^2) = (1 + c_2(K3)) \wedge 1
= 1 + c_2(K3)\;,
\ee
where we used $c_1(K3)=0$ and $c_1(T^2)=0$.
Since $c_2(K3)$ is the Euler class of $K3$ and since 
$K3$ has Euler number 24 we find
\be
C_2 \cdot P = \chi(K3)\, p^1 = 24 \,p^1\,.
\ee
So far we have discussed only topological aspects of the threefold.
But in order to get a supersymmetric state by wrapping a D4- or M5-brane
on $P$, the four-cycle must be holomorphic \cite{BecBecStr} and must
therefore define a divisor in $X$. A divisor is
an integer linear combination
of irreducible analytic hypersurfaces, where the word analytic  
emphasizes that a divisor can be locally characterized as the
zero locus of a {\em holomorphic} function.
The corresponding 
homology classes are called algebraic.
According to the Lefschetz theorem
on $(1,1)$-classes, every cohomology class in
$H^{2}(X,{\bf Z}) \cap H^{1,1}(X,{\bf R})$ is algebraic.
Here the integer valued cohomology group $H^2(X,{\bf Z})$ 
might be visualized as a lattice in the real vector space
$H^2(X,{\bf R})$, whereas $H^{1,1}(X,{\bf R})$ is a linear subspace.
If the dimension of $H^{1,1}(X,{\bf R})$ is smaller than the
dimension of $H^2(X,{\bf R})$ and if the subspace $H^{1,1}(X,{\bf R})$
is generic, then it will intersect the lattice $H^2(X,{\bf Z})$ in
no point but the origin. Therefore it is in general 
not guaranteed that algebraic classes exist.

For generic Calabi-Yau threefolds, $h_{2,0}(X)=0$, and 
therefore  $H^{1,1}(X,{\bf R})$ will coincide with $H^{2}(X,{\bf R})$,
implying that all fourth homology classes are algebraic. However for
$K3 \times T^{2}$ we have $h_{2,0}=1$
and therefore $H^{1,1}(X,{\bf R})$ is only contained in, but not
identical with $H^{2}(X,{\bf R})$. 
Obviously the problem resides
in the $K3$ factor, which has $h_{2,0}(K3) = 1$. Whether or not a $K3$
space has algebraic classes is controlled by the so-called algebraic
lattice, or Picard lattice of $K3$,
\be
\Gamma_P =H^{2}(K3,{\bf Z}) \cap H^{1,1}(K3,{\bf R}) \,.
\ee
Special $K3$ spaces, which
have a non-empty $\Gamma_P$ and therefore have holomorphic
cycles, are called algebraic. Hence we will below restrict ourselves to
these algebraic $K3$ surfaces, which are associated to subspaces of
codimension 1 in the moduli space of complex 
structures.\footnote{
  Conformal field theories describing type-II strings in the
  background of such algebraic $K3$ spaces have been discussed in
  \cite{AspMor}. 
In particular the moduli spaces of algebraic $K3$ string
backgrounds factorize into two subspaces 
containing the complex structure moduli and the
complexified K\"ahler moduli, respectively. If one
compactifies type-II strings on $K3 \times T^{2}$, then
this is the part of the moduli space that can be described
in an $N=2$ truncation of the $N=4$ theory.} 

The computation of the microscopic entropy amounts to
counting the massless excitations of a D4- or M5-brane
wrapped on $P$. In addition to being holomorphic, one has to require
that $P$ is a so-called `very ample' divisor in order to ensure
that the entropy calculation is not affected by
$\alpha^\prime$-corrections in the type-II picture and by
instanton corrections in the M-theory picture, as we will 
discuss in due
course. By definition a divisor $P$ in a space $X$
is called very ample if $X$ can be embedded into a
complex projective space ${\bf CP}^N$ such that $P$ is the
intersection of $X$ with some hyperplane in ${\bf CP}^N$. Then
Bertini's theorem implies that, generically,  
$P$ is a smooth manifold. In
our particular case the 
divisor $P$ is a compact, complex K\"ahler surface. 
On general grounds one therefore expects that the collective modes
of the wrapped brane are controlled by the Hodge numbers $h_{p,q}(P)$ 
of $P$. Therefore we will discuss how the independent Hodge numbers 
$h_{1,0}(P)$, $h_{2,0}(P)$ and $h_{1,1}(P)$
of $P$ are related to known quantities of $X$. 

Three standard invariants can
be computed using index theorems (only two of them
are independent due to the fact that a complex surface
can only have two independent Chern classes). These are the Euler number
$\chi$, the holomorphic Euler number $\chi_h$ and the Hirzebruch signature
$\sigma$, defined by
\be \begin{array}{lclcl}
\chi(P) &=& \sum_{p,q} (-1)^{p+q} \,h_{p,q}(P) &=& 2 - 4 h_{1,0}(P) 
+ 2 h_{2,0}(P) + h_{1,1}(P)\;, \\
\chi_{h}(P) &=& \sum_q (-1)^q \,h_{0,q}(P) &=& 1 - h_{1,0}(P) +
h_{2,0}(P)\;, \\ 
\sigma(P) &=& b_{2}^{+}(P) - b_{2}^{-}(P) &= & 2h_{2,0}(P) - h_{1,1}(P) +2\;,
\end{array}
\ee
where we used that $b_2^{\pm}(P)$, which denote the number of
(anti)selfdual harmonic two-forms, are equal to  
$b_2^+(P) = 2 h_{2,0}(P) + 1$ and 
$b_2^-(P) = h_{1,1}(P) - 1$ for a compact K\"ahler surface. 
Now $\chi(P)$ and $\sigma(P)$ can be computed using the following 
index theorems:
\be
\chi(P) = \int_P c_2(P)\;,\qquad 
\sigma(P) = \int_P \left( 
- \ft{2}{3} c_2(P) + \ft{1}{3} c_1(P)^2 \right) \;,
\ee
which then determines $\chi_h(P)=\ft14 (\chi + \sigma)$ as well.

To every very ample divisor $P$ in $X$
one can associate a line bundle $L$ over $X$, such 
that $P$ is the zero locus of a holomorphic section
of $L$. 
The space 
$H^{0}(X,L)$ of holomorphic sections of $L$ is 
closely related to the moduli space ${\cal M}(P)$ of
$P$, i.e. to the
space of deformations of $P$ inside $X$.
Since divisors are obtained as
zero loci of sections, multiplying a section by a non-vanishing
complex number does not change the divisor. Therefore the
moduli space of $P$ is the projective space associated
to the vector space $H^{0}(X,L)$:
\be
{\cal M}(P) = PH^{0}(X,L).
\ee
Another way of thinking about the moduli of $P$ is as
sections of the normal bundle $N_{P/X}$ of $P$,
\be
N_{P/X} = \left. {\cal T}X \right|_{P} / {\cal T}P \;,
\ee
where $\left. {\cal T}X \right|_{P}$ is the holomorphic tangent
bundle of $X$ restricted to $P$ and ${\cal T}P$ is the 
holomorphic tangent bundle of $P$. When wrapping a 
D-brane or M-brane on a holomorphic cycle, the
scalar fields of the world volume theory corresponding to 
transverse motions of $P$ inside $X$ become sections of the
normal bundle \cite{BerSadVaf}. Both descriptions of the moduli of $P$ are
equivalent because the first adjunction formula
implies that the line bundles $N_{P/X}$ and $L$ are
isomorphic.

The next step is to express the Chern classes of $P$
in terms of the Chern classes of $X$. Because the normal
bundle $N_{P/X}$ of $P$ is the quotient of the tangent bundle
of $X$ and of the tangent bundle of $P$, the  
Whitney product formula for Chern classes implies
$c(X) = c(P) \wedge c(N_{P/X})$ or
\be
1 + c_1(X) + c_2(X) + \cdots = (1 + c_1(P) + c_2(P)) \wedge
(1 + c_1(N_{P/X})). 
\ee
Since $X$ is Calabi-Yau, we have $c_1(X)=0$. Hence one gets the
relations
$c_1(P) +  c_1(N_{P/X}) = 0$ and $c_2(P) + c_1(P) \wedge c_1(N_{P/X})
= c_2(X)$, which can be combined into
\be
c_2(P) = c_2(X) + c_1(N_{P/X})^2 \;.
\ee
What remains is to express $c_1(N_{P/X})$ by a quantity defined on
$X$. To do so we recall that the first adjunction formula
implies that the normal bundle $N_{P/X}$ is isomorphic
to the line bundle $L$. In particular the first Chern classes
of both line bundles are equal, $c_{1}(N_{P/X})=c_1(L)$.
Moreover the two-form $c_{1}(L)$ is Poincar\'e dual to the
four-cycle $P$, and one can rewrite the above integrals
as integrals over $X$:
\be
\chi(P) = \int_X \Big(c_2(X) + c_1(L)^2\Big) \wedge  c_1(L) \;,\qquad 
\sigma(P) = \int_X \Big(- \ft{2}{3} c_2(X) - \ft{1}{3} c_1(L)^2\Big )\wedge
c_1(L)\;. 
\ee
Finally the integrals on the right hand side can 
be interpreted as intersection products using
\be
P^3 := P \cdot P \cdot P = \int_X c_1(L)^3\;,\qquad 
\int_P c_2(X) = \int_X c_1(L) \wedge c_2(X) = C_2 \cdot P \;,
\ee
where $C_2$ is the two-cycle Poincar\'e dual to $c_2(X)$.
With these results one can express the invariants 
in terms of intersection numbers, 
\be
\chi(P) = P^3 + C_2 \cdot P\;,\qquad 
\sigma(P) = - \ft{1}{3} P^3 - \ft{2}{3} C_2 \cdot P \;.
\ee
Now it is straightforward to obtain the
following relations for the Hodge numbers, 
\be
\begin{array}{lcl}
h_{2,0}(P) &=& \frac{1}{6} P^3 + \frac{1}{12} C_2 \cdot P + h_{1,0}(P) -1\;,\\
h_{1,1}(P) &=& \frac{2}{3} P^3 + \frac{5}{6} C_2 \cdot P 
+ 2 h_{1,0}(P)\;, \\
\end{array}
\ee
or, equivalently,
\be
\begin{array}{lcl}
b_{2}^+(P) &=& \ft{1}{3} P^3 + \ft{1}{6} C_2 \cdot P + 2 h_{1,0}(P) 
-1 \;,\\
b_{2}^{-}(P) &=& \ft{2}{3} P^3 + \ft{5}{6} C_2 \cdot P
+ 2 h_{1,0}(P) -1 \;.\\
\end{array}
\ee
What remains is to calculate $h_{1,0}(P)$. 
Here one uses again that $P$ is a
very ample divisor.
Therefore the Lefschetz hyperplane theorem implies that
$h_{1,0}(P)=h_{1,0}(X)$. Thus we have $h_{1,0}(P)=0$
when $X$ is a generic Calabi-Yau threefold and 
$h_{1,0}(P)=1$ when $X=K3 \times T^{2}$.

The final step in the approach of \cite{MSW}
is to count the massless left-moving
bosons and fermions of the wrapped M5-brane and to
express them in terms of the topological quantities
computed above. As argued in \cite{Vafa} the
analysis of the equivalent D4-D0 system should give
the same answer. We will follow \cite{MSW}.

In flat space, the massless degrees of freedom
of the M5-brane correspond to a six-dimensional
$N=(0,2)$ tensor multiplet, which contains a selfdual
antisymmetric tensor, five scalars which describe 
transverse motions and two Weyl spinors. In order to
compute the microscopic entropy one needs to dimensionally
reduce this system on the divisor $P$ and to count the
massless modes in the non-supersymmetric, left-moving
sector of the resulting two-dimensional theory.

Dimensional
reduction of the antisymmetric tensor on $P$ gives
$b_{2}^{-}(P)$ left-moving and $b_{2}^{+}(P)$ right-moving
scalars together with $b_{1}(P)=2\,h_{1,0}(P)$ gauge fields. 
Two-dimensional gauge fields do not carry dynamical
degrees of freedom themselves, but may modify the
counting for the other modes. For a generic threefold
one has $b_{1}(P)=b_{1}(X)=0$ and this problem is absent.
A full analysis would require a detailed study of the effective
$\sigma$-model describing the collective modes of a M5-brane
wrapped on the divisor $P$ in $K3 \times T^2$. This will not be
attempted here. 

Instead we will first review the counting performed in 
\cite{MSW,Vafa} and assume that there are no modifications
for $K3 \times T^{2}$. As we will see in a moment the
resulting microscopic formula contains, in the case
of $K3 \times T^{2}$, sub-subleading terms that
are puzzling from the supergravity point of view. Moreover, 
with this unmodified mode counting the right-moving
sector of the effective M5-brane theory turns out not to be supersymmetric. 
This implies that the microscopic analysis in the case of $K3 \times T^{2}$ 
must be more subtle.
We will then propose a modification of the zero-mode counting which 
is consistent with the supergravity analysis.

The M5-brane theory in flat space has
5 scalars describing the transverse motions. 
These split 
into three scalars corresponding to the position of
the brane in the three non-compact space dimensions
and two scalars corresponding to motions of the divisor
$P$ inside the Calabi-Yau threefold. Whereas the first
set just gives three scalar zero-modes, the zero-modes
associated to the second set are more subtle. One gets
one zero-mode for every independent holomorphic deformation
of $P$ inside $X$. In other words these zero-modes  
are sections of the normal bundle $N_{P/X}$. 
As reviewed above the dimension of the corresponding
moduli space is
\be
\dim_{\fscr{R}}{\cal M}(P) 
= \dim_{\fscr{R}} PH^{0}(X,L) = \dim_{\fscr{R}} H^{0}(X,L)
-2 \;.
\ee
It was emphasized in \cite{MSW} that 
this quantity is hard to compute
in general, but that there is a Riemann-Roch index theorem 
for the holomorphic
Euler number 
$\chi_h(L)=\sum_{i} (-1)^{i} \dim_{\fscr{C}} H^{i}(X,L)$ of line bundles 
$L$ over complex manifolds.
We already mentioned that the divisor $P$ is chosen to be very ample. 
Then
the Kodaira vanishing theorem together with the fact that
$X$ is a Calabi-Yau manifold 
implies that $\dim_{\fscr{C}} H^{i}(X,L)=0$ for $i>0$. 
Therefore 
$\dim_{\fscr{R}}{\cal M}(P) = 2 \chi_h (P) -2$ and hence 
\be
\dim_{\fscr{R}}{\cal M}(P) = 2 \, [h_{2,0} (P) -  h_{1,0} (P)] \;.
\eeq
This is the number of real left- and right-moving scalars
related to transverse motions of $P$ inside $X$. 
Combining it with the counting of modes descending from
the antisymmetric tensor field, one finds that the number
of left- and right-moving bosonic degrees of freedom is:
\begin{eqnarray}
N_{\mscr{bosonic}}^{\mscr{left}} &=& \dim_{\fscr{R}} {\cal M}(P)
+ 3 + b_{2}^{-}(P) =  2 h_{2,0}(P) +  h_{1,1}(P) + 2   - 2 h_{1,0}(P) 
\;,
\nonumber \\
N_{\mscr{bosonic}}^{\mscr{right}} &=& \dim_{\fscr{R}} {\cal M}(P)
+ 3 + b_{2}^{+}(P) = 4 h_{2,0}(P) +  4  - 2 h_{1,0}(P) 
\;.
\label{bosmodes}
\end{eqnarray}
The number of real left- and right-moving fermions, on the other hand, is given
by the sum of odd and even cohomology elements, respectively
\cite{Vafa}:
\beqa
N_{\mscr{fermionic}}^{\mscr{left}} &=& 4 h_{1,0}(P) \;, \nonumber\\
N_{\mscr{fermionic}}^{\mscr{right}} &=& 4 [ h_{2,0}(P) + h_{0,0} (P) ] \;.
\label{fermmodes}
\eeqa
The effective two-dimensional theory describing the collective modes of a 
BPS black hole is a $(0,4)$ supersymmetric 
sigma-model.  Here we recall that the $(0,4)$ worldsheet supersymmetry
is crucial for describing an $N=2$ BPS black hole in four-dimensional
spacetime.
Therefore, the number of right-moving
bosons and fermions has to match.
Moreover the right-moving scalars are expected to parametrize a quaternionic
manifold and therefore the number of right-moving real bosons
should be a multiple of four.   Inspection of (\ref{bosmodes})
and (\ref{fermmodes}) shows that in the case of a generic threefold, for
which
$h_{1,0}(P)=0$, 
the counting of right-moving modes is consistent with 
$(0,4)$ supersymmetry, whereas this is not the case
for $K3 \times T^2$, for which
$h_{1,0}(P)=1$.  This implies that the zero-mode counting for
$K3 \times T^2$ has to deviate from the one described above.

Using  (\ref{bosmodes}) and (\ref{fermmodes}) the central charge
of the left-moving sector is computed to be 
\be
\begin{array}{lcl}
c_{L} &=& N_{\mscr{bosonic}}^{\mscr{left}} + \frac{1}{2}
N_{\mscr{fermionic}}^{\mscr{left}} =
P^{3} + C_{2} \cdot P + 4 h_{1,0}(P) \\
 & = &  C_{ABC} \,p^A p^B p^C +
c_{2A} \,p^A + 4 h_{1,0}(P)\,.
\end{array}
\label{cl}
\ee
For the generic case, where $h_{1,0}(P)=0$, this
leads via Cardy's formula to the final result
\cite{MSW,Vafa}
\be
{\cal S}_{\mscr{micro}} = 2 \pi \sqrt{ \ft16{|q_{0}|} \Big(
C_{ABC}\, p^{A} p^{B} p^C + c_{2A}\, p^A \Big) } \;.
\label{genericentropia} 
\ee
In the case of $K3\times T^2$ the intersection
and second Chern class numbers take a simpler form. 
Using (\ref{cl}) with $h_{1,0}(P)=1$, the associated microscopic
entropy (\ref{cardmicro}) is computed to be
\be
{\cal S}_{\mscr{micro}} = 2 \pi \sqrt{ \ft16{|q_{0}|} \Big(
3 \,C_{ab}\, p^{a} p^{b} p^1 + 24 \,p^1 + 4 \Big) }
\;. \label{specialentropia} 
\ee
We note that the sub-subleading third term in this expression
is not consistent with the macroscopic computation of the entropy
based on $N=2$ supergravity.  As mentioned 
in the introduction, 
the homogeneity properties of the prepotential together with the
stabilization equations imply that the entropy should take the
form \eqn{expansion}, and hence terms containing odd powers of the charges
cannot be present in the entropy formula.

Another reason why this sub-subleading term is troublesome is that
the approaches of \cite{MSW} and \cite{Vafa} seem to give
a slightly different numerical value for this term.  
Namely, when substituting the Hodge numbers of $P$ into the entropy formula
\cite{Vafa}
\be
{\cal S}_{\mscr{micro}} = 2 \pi \sqrt{ \ft{1}{6} |q_0| ( 
b_{\mscr{even}}(P) + \ft12 b_{\mscr{odd}}(P) )} \;, 
\ee
where $b_{\mscr{even}}(P)$ and $b_{\mscr{odd}}(P)$ denote the sums
of the even and of 
the odd Betti numbers, respectively, we obtain
(\ref{specialentropia}) but 
with a 6 instead of a 4 in the last term.
Note that for generic threefolds, where
$h_{1,0}(P)=0$, both approaches 
\cite{MSW,Vafa} yield (\ref{genericentropia}).

How is the zero-mode counting for $K3 \times T^2$ to be modified
in order to remove the inconsistencies mentioned above?  
Let us recall that there are $b_1 = 2 h_{1,0} (P) =2$ nondynamical gauge fields
present.  If we assume that the zero-modes are charged and couple to 
these gauge fields, then the 
following mechanism suggests itself.  Due to gauge invariance, 
the number of left- and right-moving scalar fields is reduced by two, 
so that the number of right-moving scalar fields is indeed a multiple of four.
Due to supersymmetry this must be 
accompanied by the removal of four right-moving
fermionic real degrees of freedom.  If, in addition, we assume that the
removal of fermionic degrees of freedom is left-right symmetric,
then the actual number of left-moving fermionic degrees of freedom is zero.
The central charge in the left-moving sector
is now computed to be $c_L = P^3 + C_2 \cdot P
= 3 \,C_{ab}\, p^{a} p^{b} p^1 + 24 \,p^1$, which is odd in the charges.
The resulting microscopic entropy formula is then in full agreement
with the macroscopic computation, which we now briefly describe.

Let us analyze the same system from the 
supergravity point of view. 
The Wilsonian action controlling the vector multiplet sector
of an $N=2$ compactification and the relevant $N=2$ subsector
of an $N=4$ compactification is encoded in the
holomorphic function
\be
F(Y, \Upsilon) = - \ft{1}{6} C_{ABC} \frac{Y^{A}Y^{B}Y^{C}}{Y^{0}}
- \ft{1}{24}\ft{1}{64} c_{2A}\, \frac{Y^{A}}{Y^{0}}\, \Upsilon
= - \ft{1}{2} C_{ab} \frac{Y^1 Y^{a}Y^{b}}{Y^{0}}
- \ft{1}{64} \frac{Y^1}{Y^{0}}\, \Upsilon \;,
\label{GenPrep}
\ee
where the first formula refers to the generic case, whereas
the second formula refers to a compactification on $K3 \times T^2$.
In the latter case
$Y^{a}/Y^{0}$ are the complexified K\"ahler moduli of an algebraic
$K3$ and $Y^1/Y^{0}$ is the complexified K\"ahler modulus of $T^{2}$.

Since the function (\ref{GenPrep}) represents the leading part of the
holomorphic function in the
$\alpha'$-expansion\footnote{
  In the M-theory picture $\alpha^\prime$-corrections appear as
  instanton corrections corresponding
  to wrapped M-branes. For convenience we will use the type-IIA 
  picture in the following.},  
we briefly discuss under which conditions
a solution based on this function can be  reliable \cite{BCDKLM,MSW}. 
The issue of $\alpha'$-corrections is not quite the same for
compactifications on generic threefolds and for compactifications on
$K3 \times T^2$.  For generic
threefolds, which lead to $N=2$ compactifications, both the
prepotential $F(Y,\Upsilon=0)$ and the $C^2$-coupling function
$F_{\Upsilon}$ receive $\alpha'$-corrections. In contrast
the metric on the $N=4$ moduli space and, hence, 
the prepotential $F(Y,\Upsilon=0)$ in the $N=2$ subsector of
an $N=4$ compactification,
do not receive $\alpha'$-corrections. However the gravitational
$C^{2}$-coupling function $F_\Upsilon$ does  
get corrected \cite{HM}. 
Therefore (\ref{GenPrep}) is valid for both $N=2$ and $N=4$
compactifications in the 
limit of large K\"ahler moduli, only. In particular the 
K\"ahler moduli $\mbox{Im}(Y^A / Y^0)$ must take large positive values
at the event horizon. Since the 
K\"ahler moduli are space dependent for a generic extremal
four-dimensional black hole solution, the geometry of the internal
manifold will change accordingly over space. The values of the
moduli at the event horizon are fixed by the stabilization equations.
For a black hole with non-vanishing charges $q_0, p^A$ we have 
\cite{CDWM}
\be
\frac{Y^A}{Y^0} \Big\vert_{\mscr{horizon}} 
= i p^A \sqrt{ \frac{6 |q_0| }{ C_{BCD}\, 
p^B p^C p^D + c_{2B} \, p^B } } \;.
\label{StabEqua}
\ee
Thus, in order to guarantee 
$\mbox{Im} \,{Y^A}/{Y^0} \vert_{\mscr{horizon}} \gg 0$ we need to
impose $|q_0| \gg  p^A > 0$. 
In fact we will impose the stronger condition $|q_0|\gg p^A\gg 0$ in order
to make contact with the microscopic entropy counting. 

{From} the microscopic point of view the charges $p^A$ are the expansion
coefficients of the divisor $P$ in the homology basis,
$P=p^A \Sigma_A$. By Poincar\'e duality there is a dual $(1,1)$-form
$p^A \omega_A$, where $\omega_A$ belong to the basis of $H^2(X,{\bf Z})$
introduced above. The K\"ahler form of the internal threefold is
proportional to $\mbox{Im}(Y^A/Y^0) \omega_A$. Note that the 
stabilization equations (\ref{StabEqua}) imply that the K\"ahler form,
evaluated with the values that the K\"ahler moduli take at the horizon,
is proportional to the Poincar\'e dual of $P$. 
If all $p^A$ are positive, the
$(1,1)$-form $p^A \omega_A$ lays in the interior of the K\"aher cone
and this provides a
link between suppression of $\alpha'$ corrections and $P$ being a 
very ample divisor. Namely, by multiplying the $(1,1)$-form with a large
positive number or, equivalently, by 
taking the $p^A$ very large,  we 
can arrange that $p^A \omega_A$ is far away from the boundaries
of the K\"ahler cone. In terms of homology 
$p^A > 0$ implies that the divisor $P$ has positive intersection
numbers with all complex submanifolds of $X$. This shows that
$P$ is a so-called ample divisor according to the Nakai-Moishezon criterion.
By definition a divisor is called ample if it can be made very ample
by multiplying it with a sufficiently large positive number. Thus
$P$ is ample if $p^A > 0$ and very ample if $p^A \gg 0$. As we explained
above one has to require that $P$ is very ample in order to
be able to reliably compute the microscopic entropy.

We now turn to the macroscopic entropy of a black hole with
non-vanishing charges $q_0$ and $p^A$, 
which can 
be computed by substituting the function (\ref{GenPrep}) into the general
formula \eqn{entropiay}.  
In the case of a generic
threefold the resulting expression for the 
macroscopic entropy fully agrees with the microscopic
formula 
(\ref{genericentropia}), as we already showed in \cite{CDWM}.
For the case of a 
compactification on $K3 \times T^2$ we find
\be
{ \cal S}_{\mscr{macro}} = 2 \pi \sqrt{ \ft16{|q_{0}|} \Big( 3\, C_{ab}\, p^{a}
p^{b}\,p^1 + 24\, p^1 \Big)} \;.
\ee
This too is in full agreement with the microscopic computation provided
that the zero-mode counting for $K3 \times T^2$
is modified as described above.

\section{A non-perturbative example \label{secanom}}
\setcounter{equation}{0}

Let us now consider an extension of the  heterotic-like model of 
section \ref{treehetb}, where the term proportional to $\Upsilon$ is
replaced by a more general function of $S= -iY^1/Y^0$,
\bea
F(Y,\Upsilon) = - \frac{Y^1 Y^a \eta_{ab} 
Y^b}{Y^0} + \; F^{(1)} (S) \;  \Upsilon \;,
\label{hetprepnp}
\eea
where $\eta_{ab}$ is defined as before. In section \ref{treehetb} we
noted that the dilaton and the black hole entropy depended only on
combinations of the electric and magnetic  charges that are invariant
under classical (i.e., continuous) target-space duality transformations. For 
the
extension discussed here, target-space duality remains realized, so we will
first analyze this aspect in some more detail. Subsequently we will
consider the possible invariance under $S$-duality.

First of all, the invariance under the SO$(1,n-2)$ subgroup of the
target-space duality group is manifest, as this is an invariance of
\eqn{hetprepnp}. The transformations that
extend this subgroup to the full SO$(2,n-1)$ group of target-space
dualities depend on $2n-1$ parameters, which we write as two vectors
$a_a$ and $b^a$ and a scalar $c$ under the subgroup. The charges $M_I$
and $N^I$ transform under the corresponding infinitesimal
transformations according to  
\be
\begin{array}{rcl} 
\d M_0 &= & c\, M_0 + b^a\,M_a \,,\\ 
\d M_1 & = & -c\, M_1 -\ft12 \eta^{ab} \,a_a\,M_b \,,\\ 
\d M_a & = & a_a\,M_0 -2 \eta_{ab} \,b^a\,M_1 \,,
\end{array} 
\qquad
\begin{array}{rcl}
\d N^0 & = & -c\, N^0 - a_a\,N^a \,, \\ 
\d N^1 & = & c\, N^1  + 2\eta_{ab} \,b^a\,N^b\,, \\ 
\d N^a & = & - b^a \,N^0 + \ft12 \eta^{ab} \,a_b \,N^1 \,.
\end{array}
\ee
Using that the vectors  $(Y^0, F_1, Y^2, \dots, Y^n)$ and $ (F_0,
-Y^1, F_2, \dots, F_n)$ transform as $N^I$ and $M_I$ under
target-space duality \cite{Ceresole,DWKLL}, the corresponding
transformations of $Y^I$ and $F_I$ are given by
\be
\begin{array}{rcl}
\d Y^0 & = & -c\, Y^0 - a_a\,Y^a \,, \\ 
\d Y^1 & = & -c\, Y^1  + \ft12 \eta^{ab} \,a_a \,F_b \,, \\ 
\d Y^a & = & - b^a \,Y^0 + \ft12 \eta^{ab} \,a_b \,F_1 \,,
\end{array}
\qquad
\begin{array}{rcl} 
\d F_0 & = & c\, F_0 + b^a\,F_a \,,\\ 
\d F_1 & = & c\, F_1 +2 \eta_{ab} \,b^a\,Y^b \,,\\ 
\d F_a & = & a_a\,F_0 + 2 \eta_{ab} \,b^a\,Y^1 \,. 
\end{array} 
\ee
The crucial obervation is now that $S$ is invariant under the above
transformations of the $Y^I$, whereas these transformations induce the
correct transformations on the derivatives $F_I$ as specified
above. This means that the continuous SO$(2,n-1)$ transformations are
preserved for 
any function $F$ of the type given in \eqn{hetprepnp}. This explains
the manifest target-space duality invariance of the expressions found
in section~2.  

We now proceed and analyze the behaviour under $S$-duality. In
general it is not known which subset of the $S$-duality transformations
will be realized in $N=2$ heterotic string compactifications. But we
consider this option, first as an example to appreciate the
relevance of the non-holomorphic corrections for the entropy formula,
and secondly with an eye towards the $N=4$ theory of which the $N=2$
theory is just a truncation. 
On the electric and the magnetic charges,  the $S$-duality
transformations act according to 
\be
M_I \to \tilde M_I ={a}\, M_I - 2\, 
{b}\,\eta_{IJ} N^J \,,\qquad N^I \to
\tilde N^I =  { d}\, N^I -\ft12 \, {c}\, \eta^{IJ} M_J \, ,
\label{mntilde}
\ee
where the parametes ${a}$, $b$, $c$ and 
$d$ are integers satisfying
${ ad-bc}=1$, and where the symmetric matrix $\eta_{IJ}$ and its inverse
$\eta^{IJ}$ are defined by 
\be
\eta_{IJ} N^I N^J= N^0 N^1 + \eta_{ab} N^aN^b \,,\qquad 
\eta^{IJ} M_IM_J = 4 M_0M_1+ \eta^{ab} M_a M_b \,. 
\ee
The above expressions are equal to two of the SO$(2,n-1)$
invariants, $\ft12\langle N,N\rangle$ and $2\langle M,M\rangle$, defined in
\eqn{invcomb}. Under $S$-duality the three invariants transform according
to
\bea
\langle M,M \rangle & \to &  { a^2} \,\langle M,M \rangle 
+ { b^2} \, \langle N,N
\rangle - 2 \,{ab} \, M\cdot N \,, \nonumber \\ 
\langle N,N \rangle & \to & {c^2} \,\langle M,M \rangle 
+ {d^2} \, \langle N,N
\rangle - 2 \,{ cd} \, M\cdot N \,, \nonumber \\
M\cdot N & \to & - {ac}\,\langle M,M \rangle - {bd} \, \langle N,N
\rangle  + ({ ad + bc}) \, M\cdot N \,.
\label{mtilde}
\eea
The quantities $Y^I$ and $F_I$ transform as
\be
\begin{array}{rcl}
Y^0 &\to& \tilde Y^0= {d}\, Y^0 +{c} \,Y^1 \,, \\
Y^1 &\to& \tilde Y^1 = {a}\, Y^1 +  {b}\, Y^0 \,, \\ 
\qquad Y^a &\to& \tilde Y^a = {d} \,Y^a  -\ft12 \, {c}\, 
\eta^{ab}\,F_b \,, 
\end{array} 
\qquad
\begin{array}{rcl}  
F_0 &\to& \tilde F_0= {a}\, F_0 -{b} \,F_1 \,, \\
F_1 &\to& \tilde F_1 = {d}\, F_1 -{c}\, F_0 \,, \\
F_a &\to& \tilde F_a = {a}\, F_a - 2 \, {b}\, \eta_{ab}\,Y^b\,.
\end{array}
\ee
As a result of these transformations one easily verifies that $S$
transforms according to the well-known SL(2) formula, 
\be
S\to \tilde S =  {{a}\,S -i{b}\over i{c} \,S +{d}}\,.
\label{stilde}
\ee
However, in this case the transformations of the $F_I$ are not, in
general, correctly induced by the transformations of the $Y^I$. This
is only the case when 
\be
f(\tilde S) = (i {c} \,S + {d})^2 \, f(S)\,,
\label{holoanom}
\ee 
where $f(S) = -i\, \pa F^{(1)} (S)/\pa S$. This implies that $f(S)$
must transform under $S$-duality transformations as a modular form of
weight 2.   

In heterotic-like string compactifications, the Wilsonian coupling function
$F^{(1)} (S)$ has an expansion of the type $F^{(1)} (S) = ic_1 \, S  
+ f(q)$, where $f(q)$ denotes a series expansion in positive powers of
$q= \exp (-2 \pi S)$ and $c_1$ is the constant appearing in
\eqn{hetprep} (actually, in $N=2$ heterotic string compactifications
 the function $F$ will in general 
also receive perturbative
corrections which depend on the moduli $T^a = - i Y^a/Y^0$.
Here we do not discuss these perturbative corrections).
It is well known that there is no such holomorphic
$F^{(1)} (S)$ satisfying (\ref{holoanom}).  Thus, there is no holomorphic
Wilsonian coupling function consistent with $S$-duality.  This, however,
does not pose a problem, since in the presence of massless fields
the Wilsonian coupling functions do not, in general, exhibit 
all the invariances of the physical effective coupling functions.
In order to obtain a coupling function that is consistent with
(\ref{holoanom}), we have to give up holomorphicity,  which is a
characteristic feature of Wilsonian couplings, 
and assume that
$F^{(1)}$ depends both on $S$ and on $\bar S$. In doing so, we can
still preserve the stabilization equations (where the $F_I$ remain
given by $\pa F/\pa Y^I$) and moreover the
(classical) target-space duality invariance remains intact. A
non-holomorphic $F^{(1)}(S, {\bar S})$ which satisfies 
(\ref{holoanom}) and which in the weak coupling limit ${\rm Re} \,S
 \rightarrow \infty$ 
turns into $F^{(1)}(S, {\bar S}) \rightarrow  ic_1 \, S$ is given by 
 $F^{(1)}(S, \bar S) = -i c_1\, \ft{6 }{\pi} \left(  \log \eta^2 (S)
+ \log (S + {\bar S}) \right)$, so that $f(S, {\bar S}) =
-i \partial_{S} F^{(1)}(S, \bar S) = c_1\,\ft{3}{\pi^2} G_2(S,\bar S)$,
where $ G_2(S,\bar S) = 
G_2(S) - 2 \pi/(S + \bar S)$ and $G_2(S) = - 4 \pi 
\, \partial_S \log \eta (S) 
$ (see, e.g. \cite{HM}). We should stress
here that $F^{(1)}$ is only determined up to an anti-holomorphic
function of $\bar S$, which we suppressed in view of the fact that
it should vanish for weak coupling.  Thus, a function $F$ which is consistent
with $S$-duality is given by (\ref{hetprepnp}) with
$F^{(1)}(S)$ replaced by $F^{(1)}(S, \bar S)$.  We note that 
$F^{(1)}(S, \bar S)$ is not invariant under (\ref{stilde}), 
but rather
transforms as
$F^{(1)}(S, \bar S) \rightarrow F^{(1)}(S, \bar S) + i c_1 \ft{6}{\pi}
\log(- i{c} \bar S + d)$.

Let us now turn to the entropy calculation using the same definitions
for the stabilization equation and the entropy as before, but now with
$F^{(1)}(S)$ replaced by $F^{(1)}(S,\bar S)$. First we
determine  $|Z|^2 = p^I F_I(Y, {\bar Y}, \Upsilon) - q_I Y^I$.
By following the same steps as in section~2 we obtain
\be
|Z|^2 = ( S + {\bar S} ) \, \Big[ \ft{1}{2} \, \langle N,N \rangle
+ ( f(S,\bar S) \,\Upsilon + \mbox{ h.c.}) \Big] \;\;,
\label{znp}
\ee
which reduces to (\ref{hzz}) in the weak coupling limit ${\rm Re} \,S
\rightarrow \infty$. 
The value of the dilaton $S$ at the horizon 
is, in principle, determined in terms of the charges $M_I$ and $N^I$ carried
by the black hole through the stabilization equations
(\ref{nhstaby}). Again, following the very same steps as in section 2, we
find
\bea
S\bar S \,\langle N,N\rangle &=&  \langle M,M\rangle  -2(S + {\bar
S}) ( f(S,\bar S)\,S \,\Upsilon  + \mbox{ h.c.})\,, \nonumber\\
(S- {\bar S}) \,\langle N,N\rangle &=& 2i\, M\cdot N + 2 \,(S + {\bar S})
\,(f(S,\bar S)\,\Upsilon - \mbox{ h.c.} )\,.
\label{sn}
\eea
These two equations are combined into
\be
f(S,\bar S)\,\Upsilon = \ft{1}{4}\, \frac{1}{(S + \bar S)^2}\, (M_I - 2 i \,
\bar S\, \eta_{IK} \,N^K) \, \eta^{IJ} \, 
(M_J - 2 i \,
\bar S \,\eta_{JL} N^L) \,. \label{sn2}
\ee
Note that this equation should not be regarded as the solution for the
function $f(S,\bar S)$ but rather as an equation that 
determines $S$ in terms of the charges. It is clear that this
equation is consistent with the required $S$-duality transformation of
$f(S,\bar S)$, since  
\beqa
S + \bar S &\rightarrow& \frac{S + \bar S}{|i c S + d|^2} \;\;, \nonumber\\
M_I + 2 i \, S \,\eta_{IK} \, N^K &\rightarrow& 
\frac{ M_I +2 i \,S \,\eta_{IK} \, N^K}{
i {c} S + {d}}\,,
 \eeqa
under the transformations (\ref{mntilde}) and (\ref{stilde}).
Using \eqn{sn2} the expression for $|Z|^2$ takes a very simple form,
\be
|Z|^2 =  {(M_I + 2 i \, S \, \eta_{IK} \,N^K) \, \eta^{IJ} \, 
(M_J - 2 i \, \bar S \,\eta_{JL} N^L)\over 2(S + \bar S)} \,,
\ee
whose invariance under $S$-duality is manifest. 

Using the result for $F^{(1)}$ with the non-holomorphic corrections
included, one evaluates the expression for the entropy. As it turns
out, the result is {\it not} invariant under $S$-duality, which forces
us to include an extra term.  Thus, not only does the symplectic
vector $(Y^I, F_I (Y, \bar Y, \Upsilon))$ receive non-holomorphic corrections,
but there is an additional non-holomorphic correction to the 
coupling function of the $C_{\mu \nu \rho \sigma}^2$-term, with the result that
its effective $S$-duality invariant coupling function is given by
$F^{(1)} (S, {\bar S}) + i \, c_1 \, \ft{3}{\pi} \, 
 \log (S + {\bar S}) $.
The combined $S$- and $T$-duality invariant expression for the entropy
then reads (with $\Upsilon = -64$)   
\bea \label{entropiaynonholo}
{\cal  S}_{\rm macro} &=& \pi\, \Big[ \; |Z|^2 
 +4 \, {\rm Im} \,\Big( \Upsilon\, F^{(1)} (S, {\bar S}) + i \, c_1 \,
 \ft{3 }{\pi} \, \Upsilon  \log (S + {\bar S}) 
 \Big)\; \Big] \\
&=& { \pi\over 2} \, {(M_I + 2 i \, S \, \eta_{IK} \,N^K) \, \eta^{IJ} \, 
(M_J - 2 i \, \bar S \,\eta_{JL} N^L)\over S + \bar S} + 768\,c_1 
\, \log \Big[ (S+\bar S)\,|\eta(S)|^4\Big]  \, \,.\quad~ 
\nonumber
\eea
We should stress here that the first term does implicitly depend on
the function $f(S,\bar S)$ through the solution of \eqn{sn2} for $S$. 

The above $N=2$ example can be viewed as
describing an $N=2$ subsector of the effective Lagrangian of
heterotic string theory compactified on a six-torus. We already
discussed this truncation in section~2. Hence \eqn{entropiaynonholo}
can be promoted to an $N=4$
target-space duality invariant result. This is accomplished by simply 
replacing the SO$(2,n-1)$ invariants bilinear in the charges by the
corresponding SO$(6,22)$ ones.  
Indeed, it is known \cite{HM} 
that the effective Lagrangian of this $N=4$ theory
possesses a $C^2_{\mu \nu \rho \sigma}$-term with an effective
coupling function given by 
$F^{(1)} (S, {\bar S}) + i \, c_1 \, \ft{3}{\pi} \, 
 \log (S + {\bar S}) $.
Since the $N=4$ theory is conjectured to be invariant under $S$-duality, 
the black hole entropy in the heterotic $N=4$
theory should also be invariant under strong-weak coupling duality.
It would be interesting to make contact between the $N=4$ extension of
(\ref{entropiaynonholo}) and the entropy formula of \cite{DVV}.

The equation \eqn{sn2} for $S$ can be solved iteratively
order-by-order in $\Upsilon$. 
In the case of purely imaginary $S$, the solution remains
unmodified, that is $S = i M \cdot N/\langle N,N \rangle$.

\section{Corrections from $T$-dependent higher-derivative
interaction terms}
\setcounter{equation}{0}
So far we considered holomorphic functions $F(Y,\Upsilon)$ that
depended at most linearly on $\Upsilon$. In this section we will
study functions that depend more generally on $\Upsilon$, which we
recall, is just proportional to the square of the auxiliary tensor
field $T^{ab\,ij}$ which is the lowest component of the Weyl
multiplet. More specifically, we will consider functions
$F(Y,\Upsilon)$  associated with a type-IIA string compactification on
a Calabi--Yau threefold.  
In the limit of large K\"ahler moduli, the corresponding homogenous
holomorphic function is given by
\bea
F(Y,\Upsilon) 
= \frac{D_{ABC} Y^A Y^B Y^C}{Y^0} + d_{A}\,  \frac{Y^A}{Y^0} \; \Upsilon
+  G( Y^0, \Upsilon) \;,\; 
\label{prep2acorr}
\eea
where $D_{ABC} = - \ft{1}{6} \, C_{ABC} \,,\,
d_A = - \ft{1}{24} \, \ft{1}{64}\, c_{2A}$.  The coefficients $C_{ABC}$ 
denote the intersection numbers of the four-cycles of the Calabi-Yau
threefold, whereas the coefficients $c_{2A}$ denote its second Chern-class
numbers.  The function $G(Y^0,\Upsilon)$
is proportional to \cite{GV}
\beqa
I(\alpha) 
= \ft{\alpha^2}{4} \;  \sum_{n \in Z, n\neq 0} 
\int_{0}^{\infty}
\frac{ds}{s} \; \frac{1}{\sinh^2 (\ft{\alpha s}{2})} \;
{\rm e}^{- 2 \pi i n s} \;\;\;.
\label{ialfa}
\eeqa
Here $\alpha$ is proportional to $\sqrt{\Upsilon}/Y^0$. As we shall
argue below the 
proportionality factor relating $G(Y^0,\Upsilon)$ to $I(\alpha)$ 
is given by 
$G(Y^0,\Upsilon) = - i /(2(2\pi)^3) \; \chi \; 
(Y^0)^2 \; I(\alpha)$, where $\chi$ denotes the Euler number of the
Calabi-Yau manifold. Observe that $G$ satisfies the homogeneity
relation
\be
Y^0\,G_0 + 2 \Upsilon\,G_\Upsilon = 2 G\,, \label{homogen}
\ee
where $G_0$ and $G_\Upsilon$ denote the derivatives of $G$ with
respect to $Y^0$ and $\Upsilon$, respectively.

An explicit evaluation of the integral (\ref{ialfa}) has
been given in two regimes \cite{GV}, namely one in which 
$\alpha$ is taken to be small 
and the other in which $\alpha$ is taken to be large.
When computing (\ref{ialfa}) a regularisation prescription
needs to be adopted.  In the case of small $\alpha$, the regularised
expression (\ref{ialfa}) reads \cite{GV,MM}
\beqa
I(\alpha)_{\rm reg}  
= \sum_{g=0, 2, 3, \dots} 
{b}_g \;  \alpha^{2g} \quad, \qquad
{b}_g =   (-)^{g} \; 2 (2g-1) 
\frac{\zeta (2g) \,\zeta(3- 2g)}{( 2 \pi)^{2g}}
 \;\;\;.
\label{expi}
\eeqa
The coefficient ${b}_0$, 
in particular, is given by 
${b}_0 = - \zeta (3)$, which shows that the 
proportionality factor relating 
$G(Y^0, \Upsilon)$ to $I (\alpha)_{{\rm reg}}$ must be
the one mentioned above, since the function $F$ is known to contain a
term $i/(2(2\pi)^3) \; \zeta (3) \; \chi 
\;  (Y^0)^2$ \cite{HKTY}. We note that, for small $\alpha$, 
 the function $G(Y^0, \Upsilon)$ is not actually fully captured by (\ref{expi})
since there are additional contributions to it which are 
not analytic near $\alpha=0$ \cite{GV}.

In a number of cases one can perform explicit calculations and
determine the value of the scalar fields $Y^I$ (from (\ref{staby}))
and of the entropy (from (\ref{entropiay})) in the presence of a
function $G(Y^0,\Upsilon)$. Below we will evaluate the case with
$p^0=0$ (so that $Y^0$ is real), the case of axion-free black holes and the
case where $Y^0$ is 
imaginary and thus equal to $\ft12 i p^0$. To facilitate this
discussion we first collect a number of useful formulae. 

First of all, the scalars $Y^A$ do not appear in the function $G$,
which implies that 
\be
F_A= {3\, D_{ABC}Y^BY^C + d_A \,\Upsilon\over Y^0}
\ee
remains unchanged. Therefore the stabilization equations for $q_A$ are 
universal (since 
these equations are only valid at the horizon, we have $\Upsilon = -64$), 
\be
q_A= {1\over |Y^0|^2} \Big[ - d_A \,p^0\,\Upsilon  - 3i D_{ABC}( Y^BY^C\bar 
Y^0  -\bar Y^B\bar Y^C Y^0)\Big] \,. 
\label{qAstab}
\ee
The stabilization equation for $q_0$ does depend on the function $G$
and reads
\be
q_0= i\, {D_{ABC}Y^AY^BY^C + d_AY^A \,\Upsilon\over
(Y^0)^2} -i\, {D_{ABC}\bar Y^A\bar Y^B\bar Y^C + d_A\bar Y^A \,\bar
\Upsilon\over (\bar Y^0)^2}  -i (G_0-\bar G_0) \,. \label{q0stab}
\ee
Furthermore, the expression for $|Z|^2 = p^I\,F_I - q_I\,Y^I$ can be
written as follows,
\bea
|Z|^2 &=&  i \,D_{ABC} \Big[ {3\,Y^AY^B\bar Y^C\over Y^0} -{3\,\bar
Y^A\bar Y^B Y^C\over \bar Y^0} - 
{Y^AY^BY^C \bar Y^0\over (Y^0)^2} + {\bar Y^A\bar Y^B\bar Y^C
Y^0\over (\bar Y^0)^2} \Big] \nonumber \\
&& + i d_A\Big [  {\bar Y^A \,\Upsilon \over Y^0} 
- {Y^A \,\bar \Upsilon \over \bar Y^0} 
- {Y^A \bar Y^0 \,\Upsilon \over (Y^0)^2} 
+ {\bar Y^A Y^0\,\bar \Upsilon \over (\bar Y^0)^2}\Big]
\nonumber \\[2mm]
&& + \ft12 i (Y^0+ \bar Y^0)(G_0-\bar G_0) +  \ft12 p^0(G_0+\bar
G_0)   \,. \label{ZZZ}
\eea
The entropy is given by 
\be
{\cal S}_{\rm macro} 
= \pi\,\Big[ |Z|^2 - 2id_A\, \Big({Y^A\over Y^0} \, \Upsilon
-{\bar Y^A\over \bar Y^0} \, {\bar \Upsilon} \,\Big) -2i (\Upsilon
\,G_\Upsilon - {\bar \Upsilon} \, \bar G_{\Upsilon}) \, \Big] \,.
\label{entrog}
\ee 
We now turn to the three separate cases. 

\subsection{Black holes with $p^0 =0$}
Let us compute the entropy for a class of type-IIA black holes
with $p^0 =0$. This implies immediately that $Y^0= \bar Y^0$. The
equation \eqn{qAstab} then simplifies and one establishes directly that  
\be
Y^A = \ft{1}{6} Y^0 D^{AB} q_B +
\ft{1}{2} i p^A \,,\qquad \mbox{where} \quad  D_{AB} \equiv D_{ABC}
p^C \,,\quad  D_{AB} D^{BC} = \delta_A^C\,,
\ee
where we recall that $D_{ABC}$ and $d_{A}$ are related to the
four-cycle intersection numbers and the second Chern-class
numbers of an underlying Calabi-Yau manifold. 
Similarly, one obtains from \eqn{q0stab},  
\be
4\, (Y^0)^2 = {D_{ABC}\,p^Ap^Bp^C - 4d_Ap^A\,\Upsilon  \over \hat q_0 +i
(G_0-\bar G_0) }\,,
\label{stabyg}
\ee
where $\hat q_0 \equiv q_0 + \ft1{12} D^{AB}\,q_Aq_B$. Furthermore,
from \eqn{ZZZ} we determine
\be
|Z|^2 = - {D_{ABC}\,p^Ap^Bp^C - 2d_Ap^A\,\Upsilon \over Y^0}  +
iY^0\,(G_0-\bar G_0)\,.  
\ee
Combining these results leads to the following equation for the
entropy, 
\be
{\cal S}_{\rm macro} = - 4\pi \, Y^0\,\hat q_0 - i \pi \,( 3Y^0 \,G_0 + 2
\Upsilon\,G_\Upsilon - \mbox{ h.c.} )\,.
\label{ent2g}
\ee
For vanishing $G$ this result agrees with that given in 
\cite{CDWM}, which was consistent with the results of state counting
presented in \cite{MSW,Vafa} in appropriate limits. 

Equation (\ref{stabyg}) can be solved iteratively for $Y^0$ in terms
of the charges, as follows.  Let us denote the value of $Y^0$ by $y^0$
when $G$ is switched off.  We take the magnetic charges $p^A$ to be positive
and ${\hat q}_0 <0$, so that $(y^0)^2 >0$, as can be seen from (\ref{stabyg}).
In a regime where $| {\rm Im} \, G_0 (y^0, \Upsilon)|
\ll |{\hat q}_0|$, (\ref{stabyg}) can, to first approximation, be written as
\beqa
Y^0 = y^0 \Big( 1 + \ft{1}{2} \, \frac{i (G_0 (y^0,\Upsilon) - {\bar G}_0
({\bar y}^0, {\bar \Upsilon}))}{|{\hat q}_0|} + \cdots  \Big) \;.
\label{yyfirst}
\eeqa
Inserting (\ref{yyfirst}) into (\ref{ent2g}) then yields, to first
approximation,
\bea
{\cal S}_{\rm macro} =
2 \pi \sqrt{\ft{1}{6}|{\hat q}_0| (C_{ABC} \,p^A p^B p^C + c_{2A}\, p^A)} 
- 2 \pi i ( G(y^0, \Upsilon) - {\bar G}
({\bar y}^0, {\bar \Upsilon}))  + \cdots \;\;\;,
\eea
where we used the homogeneity property (\ref{homogen}) for $G$ and
expressed the result in terms of the intersection numbers and second
Chern-class numbers.

\subsection{Axion-free black holes}
Now let us consider axion-free black holes, which are characterized by
the fact that the moduli $z^A= Y^A/Y^0$ are imaginary. Using $\bar Y^0
Y^A +Y^0 \bar Y^A=0$, it readily follows that 
\be
Y^A = i p^A\,{Y^0\over \lambda}\,,\qquad \mbox{where}\quad Y^0 = \ft12
( \lambda + ip^0)\,. 
\ee
The charges $q_A$ follow directly from \eqn{qAstab},
\be
q_A = -3 \,D_{ABC}\,p^Bp^C\,{p^0\over \lambda^2} - 4 d_A\,
{p^0\,\Upsilon\over  \lambda^2 + (p^0)^2}\,.
\ee
This equation implies that the charges are tightly constrained. It
also implies a quadratic equation for $\lambda^2$. Its solution thus
fixes all the moduli in terms of the charges. The stabilization
equation \eqn{q0stab} for $q_0$ shows that $q_0$ is not independent,
but given by 
\be
q_Ap^A + 3 q_0p^0 = - d_Ap^A\,\Upsilon\,{16\,p^0\over \lambda^2 + (p^0)^2} 
- 3 i p^0(G_0-\bar G_0)\,. 
\ee
In this case \eqn{ZZZ} yields
\bea
|Z|^2 &=& -2 \,D_{ABC}\,p^Ap^Bp^C {\lambda^2 + (p^0)^2\over \lambda^3} +
4\, d_Ap^A \,\Upsilon {\lambda^2 - (p^0)^2\over \lambda( \lambda^2 +
(p^0)^2) } \nonumber \\
&& + \ft12 i\lambda (G_0-\bar G_0) +  \ft12 p^0 (G_0+\bar G_0) 
\,. 
\eea
The entropy can be written as follows (with $\Upsilon = -64$)
\bea
{\cal S}_{\rm macro} 
&=& \pi\,\Big[ -2 \,D_{ABC}\,p^Ap^Bp^C {\lambda^2 + (p^0)^2\over
\lambda^3} + 8\, d_Ap^A \,\Upsilon {\lambda\over  \lambda^2 +
(p^0)^2 } \nonumber \\
&&\hspace{5mm}  + \ft12 i\lambda (G_0-\bar G_0) +  \ft12 p^0 (G_0+\bar
G_0)   -2i \Upsilon (
\,G_\Upsilon - \,\bar G_{\Upsilon})  \Big] \,. 
\eea
For $G = i c (Y^0)^2$ and $d_A=0$ this result coincides with that
obtained in \cite{BCDKLM}.

\subsection{Black holes with ${\rm Re} \,Y^0 =0$}
Finally, we consider black holes for which ${\rm Re}\, Y^0 =0$ and, hence, 
$Y^0 = \ft{1}{2}i p^0$. The stabilization equations \eqn{qAstab} and
\eqn{q0stab} yield the following two conditions
\bea
p^0 \,q_A &=&  - 6 \,D_{ABC}(Y^B Y^C+\bar Y^B\bar
Y^C) - 4d_A\,\Upsilon \,,\nonumber \\
(p^0)^2\,q_0 &=& 4\, D_{ABC}\,p^A(Y^B Y^C+Y^B \bar Y^C+ \bar Y^B\bar
Y^C) + 4d_Ap^A \,\Upsilon -i (p^0)^2 \,(G_0-\bar G_0)\,. \quad~
\label{stabrmy}
\eea
Just as in the previous subsection the charge $q_0$ is not independent
and is constrained by 
\be
p^0\, p^I\,q_I =  2 \,D_{ABC}\,p^Ap^Bp^C -i (p^0)^2 \,(G_0-\bar
G_0)\,. 
\label{charconstr}
\ee
The equations \eqn{stabrmy} are quadratic in terms of the scalars $Y^A$
which can therefore be determined in terms of the charges. However, we
do not wish to pursue this in full generality.  Below we will determine the
value of the $Y^A$ for type-II models with a dual heterotic description.

{From} \eqn{ZZZ} and (\ref{entrog}) we obtain
\beqa
|Z|^2 &=& {2\over p^0} D_{ABC}(Y+\bar Y)^A (Y+\bar Y)^B(Y+\bar Y)^C +
{4\over p^0} d_A(Y+\bar Y)^A \,\Upsilon
 + \ft12 p^0 (G_0+\bar G_0)\,, \nonumber\\
{\cal S}_{\rm macro} 
&=& {2\pi \over p^0}\, D_{ABC}(Y+\bar Y)^A (Y+\bar Y)^B(Y+\bar
Y)^C  -2i\pi (G-\bar G) \,,
\eeqa
where we made use of the homogeneity property \eqn{homogen} for $G$. 

For type-II models with a dual heterotic description, we can use the
first equation in (\ref{stabrmy}) to determine the the $Y^A$, along
the same lines as in section~2. We start from 
\be
D_{ABC}\, Y^A Y^B Y^C = - Y^1 Y^a \eta_{ab} Y^b \;\;, \qquad 
d_{A} Y^A = c_1 Y^1 \,,
\ee
so that $D_{1ab} = -\ft13 \eta_{ab}$. Then  the
first equation in (\ref{stabrmy}) yields 
\bea
Y^a&=& {1\over Y^1+\bar Y^1} \,\Big[ \ft14 p^0\,\eta^{ab}q_b + i \bar
Y^1\,p^a\Big]\,, \nonumber\\
4c_1\,\Upsilon + p^0q_1 + p^a\eta_{ab}p^b &=& {1\over (Y^1+\bar Y^1)^2}
\, \Big[\ft14 (p^0)^2 \,q_a\eta^{ab}q_b + (p^1)^2 p^a\eta_{ab}p^b + p^0
p^1\,p^aq_a \Big] \,.\quad~
\eea
Substituting this into the above entropy formula, we obtain
 \bea
{\cal S}_{\rm macro} &=& -{2\pi\over p^0(Y^1+\bar Y^1)} \,\Big[\ft14 (p^0)^2
\,q_a\eta^{ab}q_b + (p^1)^2 p^a\eta_{ab}p^b + p^0 p^1\,p^aq_a \Big]
-2i\pi (G-\bar G) \nonumber\\ 
&=&  -{\pi\over p^0} 
\sqrt{\Big(
(p^0)^2 q_a \eta^{ab} q_b + 4 (p^1)^2 p^a \eta_{ab} p^b + 4 p^0 p^1 q_a p^a 
\Big)
\Big( 
q_1 p^0 + p^a \eta_{ab} p^b + 4 c_1  \Upsilon
\Big)}  \nonumber\\
&& -2i\pi (G-\bar G) \,, 
\eea
We thus see that we have to choose $p^0 < 0$.
This expression can be rewritten as follows 
in terms of the heterotic electric and magnetic
charges $M_I$ and $N^I$ given in section \ref{treehetb},
\begin{eqnarray}
{\cal S}_{\rm macro}
&=& -{\pi\over p^0}  \sqrt{ \left( (p^0)^2 \la M, M \ra + (p^1)^2
\la N,N \ra + 2 p^0 {p^1} M \cdot N \right) 
\left( \la N,N \ra + 8 c_1 \Upsilon \right) } \nonumber \\
 & & - 2 i \pi (G - \bar{G} ) \,. \label{hetentg}
\end{eqnarray}
Observe that the charges are subject to the constraint
(\ref{charconstr}), which in the  
case at hand reads 
$p^0 M \cdot N + p^1 \langle N,N \rangle = 
-i (p^0)^2 \,(G_0-\bar G_0)$. 
Substituting this into (\ref{hetentg}) yields
\begin{eqnarray}
{\cal S}_{\rm macro} &=& \pi \sqrt{ \la M, M \ra \la N,N \ra - (M\cdot N)^2
- (p^0)^2 (G_0 - {\bar G}_0)^2 } \;
\sqrt{ 1 -  \frac{512 \, c_1 }{\la N,N \ra} } \nonumber \\ 
 & & - 2 i \pi (G - {\bar G}) \;, 
\end{eqnarray}
where we also used $\Upsilon = -64$. This expression reduces 
to (\ref{entrofin}) in the case of $G=0$.

{\large \bf Acknowledgements}\\[1mm]
\noindent
We thank K. Behrndt, R. Dijkgraaf, D. L\"ust and E. Verlinde for valuable
discussions. We also thank B. Hunt for 
clarifying discussions on the mathematical aspects of
section 3. 
G.L.C. thanks the CBPF in Rio de Janeiro  for
kind hospitality during the final stages of this work. B.d.W thanks
the Alexander von Humboldt-Stiftung for supporting his stay at the
AEI as part of the Humboldt Award Program. 

{\large \bf Note Added}\\[1mm]
\noindent
It should be noted that the revised counting of micro-states proposed
in section 3 yields results that are consistent with anomaly inflow
arguments \cite{HMM}.  We would like to thank J. A. Harvey and F. Larsen for
discussions concerning this point.


\end{document}